\documentclass[12pt]{iopart}
\usepackage{graphicx}
\usepackage{epsfig}
\usepackage{dcolumn}
\usepackage{bm}

\def\lessim{\lower.5ex\hbox{$\; \buildrel < \over \sim \;$}}
\begin{document} \hbadness=10000
\title[Non-Equilibrium Heavy Flavored Hadrons from Strangeness-Rich QGP]%
{Non-Equilibrium Heavy Flavored Hadron Yields\\ from Chemical Equilibrium Strangeness-Rich QGP}
 \author{Inga~Kuznetsova  and Johann Rafelski}
\address{Department of Physics, University of Arizona, Tucson,
Arizona, 85721, USA}

\begin{abstract}
The yields of heavy flavored hadrons emitted from strangeness-rich QGP are
evaluated within chemical non-equilibrium statistical hadronization
model,  conserving strangeness, charm, and entropy yields at hadronization.
\end{abstract}

%\pacs{25.75.Nq, 12.38.Mh, 25.75.-q, 24.10.Pa}
%25.75.Nq Quark deconfinement,
%         quark-gluon plasma production and phase transitions in relativistic
%         heavy ion collisions
%25.75.-q Relativistic heavy-ion collisions
%24.10.Pa Thermal and statistical models
%12.38.Mh Quark-gluon plasma  in quantum chromodynamics
% \maketitle
%%%%%%%%%%%%%%%%%%%%%%%%%%%%%%%
\section{Introduction}
A relatively large number of hadrons containing charm and bottom
quarks are expected to be produced in heavy ion (AA) collisions at
the Large Hadrons Collider (LHC). We are interested in how the high
strangeness yield influences heavy and multi-heavy hadron yields
(containing more than one heavy quark)~\cite{Kuznetsova:2006bh}. Our
work is differing from other recent
studies~\cite{Andronic:2006ky,Becattini:2005hb}, since these assume
that the hadron yields after hadronization are in chemical
equilibrium. However, we form the charm $c$ (and bottom $b$) hadron
yields in the statistical hadronization approach   based on a given
abundance of $u,d,s$ quarks fixed by the bulk properties of a
chemically equilibrated QGP phase. Since at LHC, and also RHIC,
conservation of baryon number is not a vital element of the
analysis, only entropy is  preserved in the following, considering
$u,d, \bar u, \bar d$ yields.

Our approach is justified by the expectation that, in a fast
break-up of the QGP formed at RHIC and LHC, the phase entropy and
strangeness will be nearly conserved during the process of
hadronization. If and when the volume doesn't change appreciably
during hadronization, the entropy conservation determines values of
hadron yields  after hadronization. For QGP in chemical equilibrium,
this implies that the final hadron yields are in general not in a
chemical equilibrium. In this work, we will show how this influences
the relative yields of heavy flavored particles in the final state.
We will investigate in quantitative terms how such  chemical
non-equilibrium yields, in the conditions we explore  well above the
chemical equilibrium abundance,  influence the expected yields of
single, and multi-heavy flavor hadrons. We show that there are
profound differences in hadron yields between the assumed hadron
equilibrium and our approach.

%%%%%%%%%%%%%%%%%%%%%%%%%%%%%%%%%%%%%%%%%%%%%%%%%%%%%%%%%%%%%%%%
\section{Statistical hadronization model with conserved yields}
%%%%%%%%%%%%%%%%%%%%%%%%%%%%%%%%%%%%%%%%%%%%%%%%%%%%%%%%%%%%%%%%
During a fast transition between QGP (Q) and HG (H) phases strange $i=s$ and heavier
$i=c$ and $i=b$ quark flavor yields are preserved, as is the specific, per rapidity,
hadronization volume, and in scaling limit, we also preserve
the entropy per unit of rapidity~\cite{Bjorken:1982qr}:
\begin{equation} \label{Sflcons}%\label{Scons}
\frac{dN^\mathrm{H}_i}{dy}=\frac{dN^\mathrm{Q}_i}{dy},
\qquad
\frac{dS^\mathrm{H}}{dy}=\frac{dS^\mathrm{Q} }{dy};
\qquad
\frac{dV^\mathrm{Q}}{dy}=\frac{dV^\mathrm{H}}{dy}.
\end{equation}
From now on we omit the upper index H. All parameters are for hadron
side, unless there is an upper index Q.

The number of particles of type `$i$' with mass $m_i$ per unit of rapidity is,
in our approach, given by:
\begin{equation}
\frac{dN_i}{dy}=\Upsilon_i \frac{dV}{dy}n_i^{\rm eq}.  \label{dist}
\end{equation}
Here, $\Upsilon_i=\lambda_i\gamma_i$ is the particle $i$ yield
fugacity, $dV/dy$ is the system volume associated with the unit of
rapidity, and $n_i^{\rm eq}$ is a Boltzmann particles phase space
density,
\begin{eqnarray}
n_i^{\rm eq} =g_i\int\frac{d^3p}{(2\pi)^3} \exp(-\sqrt{p^2+m_i^2}/T)
         =g_i \frac{T m_i^2}{2\pi^2} K_2(m_i/T), \label{distapr}
\end{eqnarray}
 where
$g_i$ is the degeneracy factor, $T$ is the temperature and $p$ is the momentum.

The particle fugacity ${\lambda_i}=e^{\mu_i/T}$ is associated with a
conserved quantum number,   of strangeness, charm, bottom,  and
baryon number: $i=s,c,b$. Antiparticles have negative chemical
potential and thus $\bar\lambda=\lambda^{-1}$. The ${\lambda}$
evolution during the reaction process is related to the changes in
densities due to dynamics such as expansion. The phase space
occupancy ${\gamma_i}$ is the same for particles and antiparticles.
Its value can change as a function of time even if the system does
not expand, for it describes buildup of the particular particle
species. For this reason ${\gamma_i}$ very often changes rapidly,
while ${\lambda}$ is more constant. It is ${\gamma_i}$ which is most
sensitive to the time history of the reaction, except if full
chemical equilibrium can be established, in which case
${\gamma_i}\to 1$.

\begin{figure}% [!b]
\centering
\includegraphics[width=7cm,height=7cm]{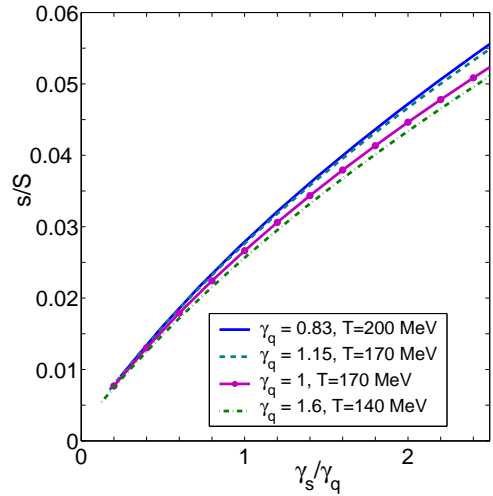}
\caption{(color on line) \small{Strangeness to entropy ratio, $s/S$,
as a function of $\gamma_s/\gamma_q$.
 (solid line, blue) for $T=200$ MeV, $S^H=S^Q\to \gamma_q=0.083$;
(dashed line, blue) for $T=170$ MeV, $S^H=S^Q\to \gamma_q=1.15$;
 (dash-dotted line, green) for  $T=140$ MeV, $S^H=S^Q \to \gamma_q=1.6$;
(dot marked solid, violet) for $\gamma_q = 1, T=170$.}}\label{sSrg}
\end{figure}

The yields of  hadrons after hadronization are given by
Eq.\,(\ref{dist}), the unknown phase occupancies on hadron side for
strangeness $\gamma_s$ and heavy quarks $\gamma_{c(b)}$ can be
determined comparing quark yields in $Q,H$ phases. The
$\gamma_s/\gamma_q$ ratio depends mostly on strangeness to entropy
ratio $s/S$ after hadronization, see fig.{\ref{sSrg}}, obtained
using SHAREv.2.0~\cite{Torrieri:2006xi}. For LHC the expected ratio
$s/S=0.038$~\cite{Letessier:2006wn} at $T=140$--180 MeV which
implies in the hadron phase
${\gamma_s}/{\gamma_q}=1.8$--2~\cite{Rafelski:2005jc}.

%%%%%%%%%%%%%%%%%%%%%%%%%%%%%%%%%%%%%%%%%%%%%%%%%%%%%%%%%%%%%%%
\section{Yields of heavy flavored hadrons}\label{heavyFlSec}
%%%%%%%%%%%%%%%%%%%%%%%%%%%%%%%%%%%%%%%%%%%%%%%%%%%%%%%%%%%%%%%

%%%%%%%%%%%%%%%%%%%%%%%%%%%%%%%%%%%%%%%%%%%%%%%%%%%%%%%%%%%%%%%
\subsection{D, Ds, B, Bs meson yields}\label{cbMesYielSec}
%%%%%%%%%%%%%%%%%%%%%%%%%%%%%%%%%%%%%%%%%%%%%%%%%%%%%%%%%%%%%%%

%%%%%%%%%%%%%%%%%%%%%%%%%%%%%%%%%%%%%%%%%%%%%%%Fig 1

\begin{figure}
\centering
\includegraphics[width=7.8cm,height=7.2cm]{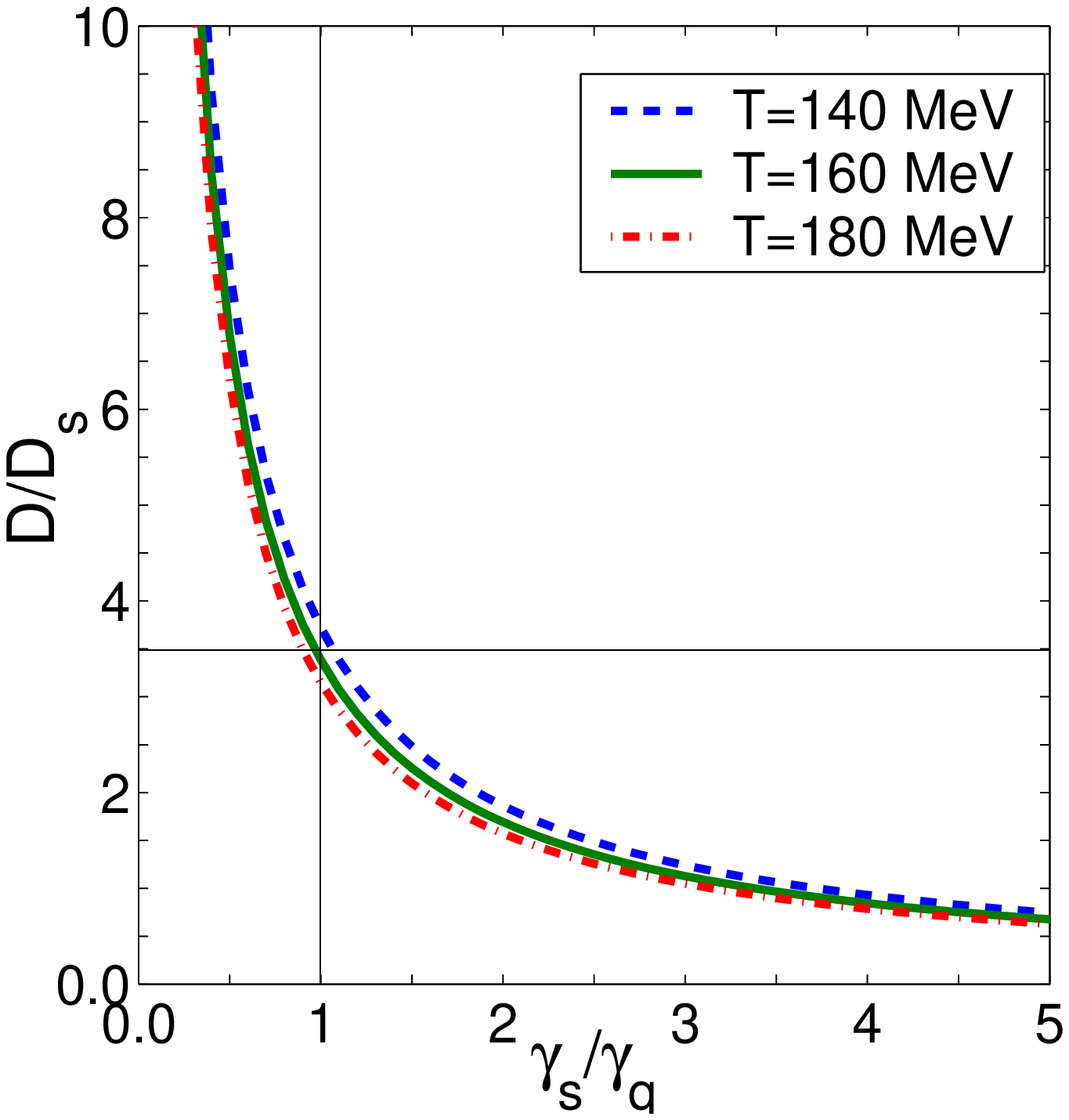}\hspace{-0.8cm}
\includegraphics[width=7.2cm,height=7.2cm]{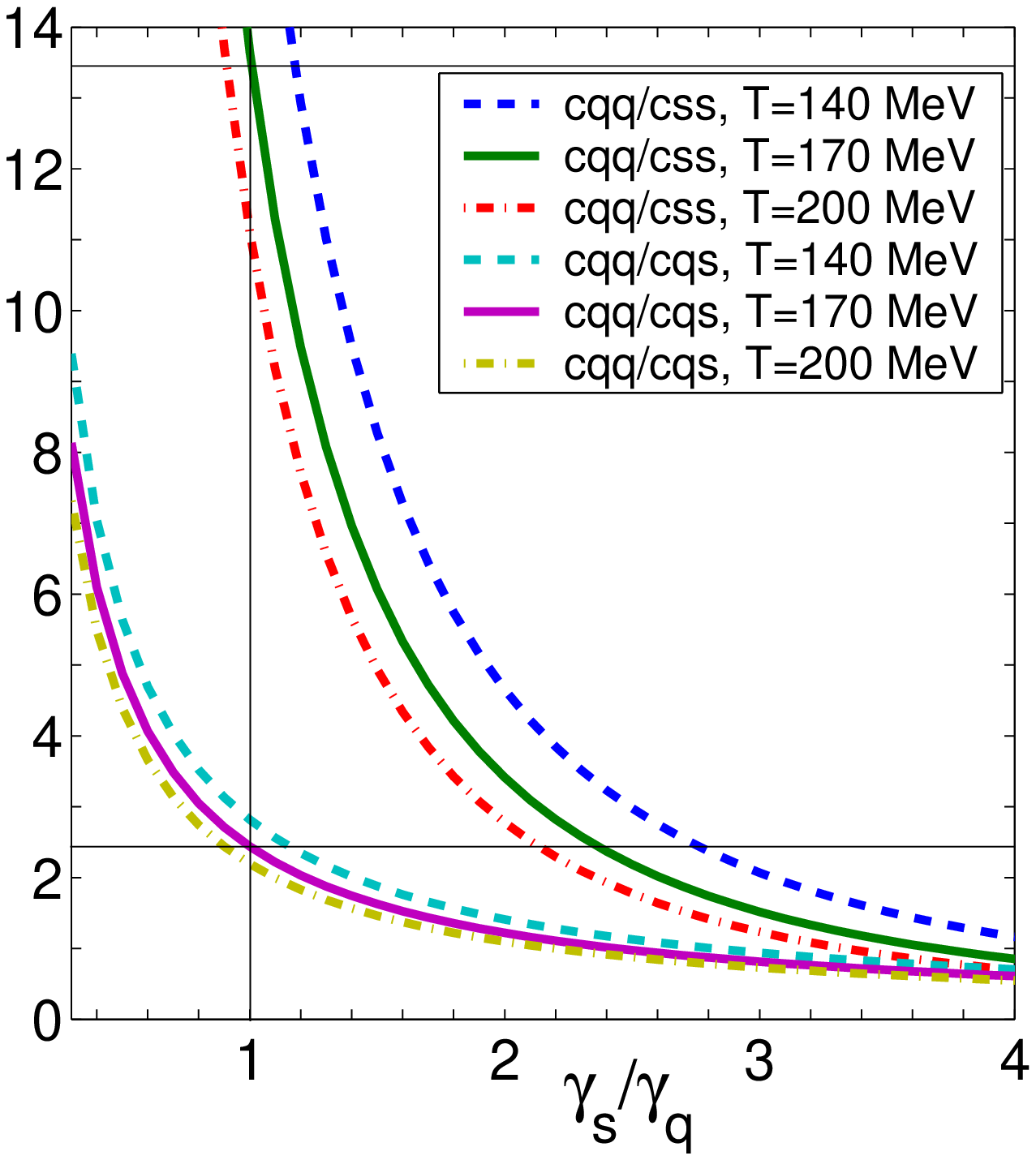}
\caption{(color on line) \small{Left: $D/D_s$ ratio as a function of
$\gamma^\mathrm{H}_s/\gamma^\mathrm{H}_q$ for $T = 140$ MeV (blue,
dashed line), $T = 160$ MeV (green, solid line) and $T= 180$ MeV
(red, dash-dot line)}. Right: The ratios $cqq/cqs=(\Lambda_c+\Sigma_c)/\Xi_c$ (lower lines)
and $cqq/css=(\Lambda_c+\Sigma_c)/\Omega_c$ (upper lines)
for $T=200 $ MeV (dash-dot line), $T=170$ MeV (solid line) and
$T=140$ MeV (dashed line).} \label{rDsDg}\label{barratio}\end{figure}

The predicted $D$ and $D_s$ mesons yield ratio $D/D_s$ is shown in figure~\ref{rDsDg} on left,
as a function of ratio ${\gamma_s}/{\gamma_q}$  for $T=140, 160, 180$ MeV.
A deviation of ${\gamma_s}/{\gamma_q}$ from unity (from chemical equilibrium)
leads to a noticeable change in the ratio $D/D_s$, which  is proportional
 to the inverse of ${\gamma_s}/{\gamma_q}$ ratio. Ratio ${\gamma_s}/{\gamma_q}$
grows with growth of strangeness to entropy ratio $s/S$.
For LHC, $s/S$ may reach a high value, and if $s/S=0.04$
then the ratio ${\gamma_s}/{\gamma_q}$ is increased by about
a factor 2   and   $D/D_s$ ratio is decreased  by a factor 2 compared to
the chemical equilibrium model. For $B$, $B_s$ mesons the results are the same
as for $D$, $D_s$ mesons.

%%%%%%%%%%%%%%%%%%%%%%%%%%%%%%%%%%%%
%%%%%%%%%%%%%%%%%%%%%%%%%%
%%%%%%%%%%%%%%%%%%%%%%%%

%%%%%%%%%%%%%%%%%%%%%%%%%%%%%%%%%%%%%%%%%%%%%
\subsection{Heavy baryon yields}\label{BarYieSec}
%%%%%%%%%%%%%%%%%%%%%%%%%%%%%%%%%%%%%%%%%%%%%

On the right, in figure~\ref{barratio}, we show ratios
$cqq/cqs=(\Lambda_c+\Sigma_c)/\Xi_c$ and
$cqq/css=(\Lambda_c+\Sigma_c)/\Omega_c$ as a function of
$\gamma_s/\gamma_q$ for $T=200$ MeV (dash-dot line), $T=170$ MeV
(solid line) and $T=140$ MeV (dashed line). The ratio $cqq/cqs$ is
inversely proportional to ${\gamma_s}/{\gamma_q}$ ratio, similar as
for $D_s/D$ ratio. However, $cqq/css \propto
(\gamma_s/\gamma_q)^{-2}$, further enhancing the effect of chemical
non-equilibrium in QGP hadronization, with multistrange-charmed baryons produced
at similar yield as nonstrange-charmed baryons.

%%%%%%%%%%%%%%%%%%%%%%%%%%%%%%%%%%%%%%%%%%%
\subsection{Yields of hadrons with two heavy quarks}\label{MultiSec}
%%%%%%%%%%%%%%%%%%%%%%%%%%%%%%%%%%%%%%%%%%%
The yields of hadrons with several heavy quarks somewhat depend on
hadronization condition, in particular on the
reaction volume and on the total assumed charm yields. If we
normalized yield of, for example, hidden charm mesons on $N_c^2$
their yield only slightly depends on total charm multiplicity. The
yields of hadrons  with two heavy quarks are approximately
proportional to $1/(dV/dy)$~\cite{Thews:2000rj}, because
$\gamma^H_{c}$ for heavy quarks is proportional to $1/(dV/dy)$~\cite{Thews:2000rj}, and
only one power is cancelled by the proportionality to volume. The
results we present are based on QGP reference state with
$dV/dy=800\,{\mathrm{fm}}^{3}$ at $T=200\,{\mathrm{MeV}}$, for $s/S
= 0.04$ and $dV/dy=600\,{\mathrm{fm}}^{3}$ at
$T=200\,{\mathrm{MeV}}$, for $s/S = 0.03$. Volume expands with drop
in temperature preserving $dS/dy$. This corresponds considering
entropy to a total final hadron particles multiplicity of about 5000
for $s/S = 0.04$.

\begin{figure}[!t]
\centering
\includegraphics[width=14.6cm,height=8.5cm]{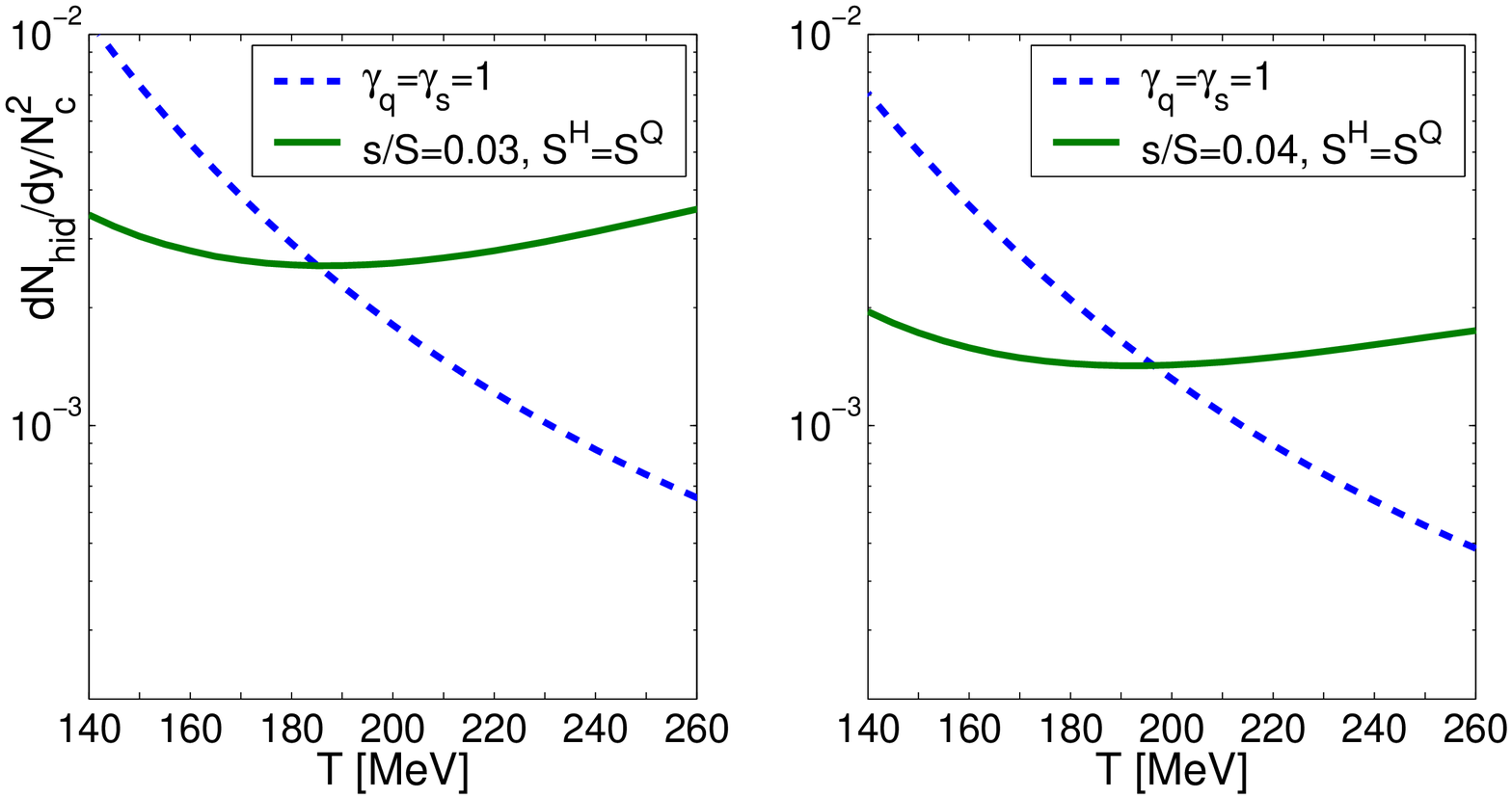}
\caption{\small{$c\bar{c}/N^2_c$ relative yields as a function of
hadronization temperature T. Dashed lines are for chemical equilibrium,
solid lines for prescribed $s/S$ in QGP. Left panel:
$dV/dy=600$ $\mathrm{fm^{3}}$ at $T=200$ MeV, with  $s/S=0.03$  for
the nonequilibrium curve. Right panel: $dV/dy=800\,\mathrm{fm^{3}}$
at $T=200$ MeV with $s/S=0.04$.}} \label{cc}
\end{figure}

In figure~\ref{cc} we show the yield of hidden charm mesons
$c\bar{c}$ (sum over all states of $c\bar{c}$)  normalized by the square
of charm multiplicity per rapidity $N_c^2$ ($N_c=10$ for LHC and $N_c=3$ for RHIC) as a
function of hadronization temperature $T$. We consider the cases $s/S=0.03$
and $s/S=0.04$ for LHC. We see that while the chemical equilibrium
model predicts a yield falling with hadronization temperature $T$,  the fixed $s/S$ yield
remains roughly independent of $T$. Remarkably, we see comparing right and left sides
of figure~\ref{cc} that in our model  the expected
yield of $c\bar{c}$ mesons decreases with increasing specific strangeness and light quarks yields. The multiplicity of light quarks is also larger than in chemical equilibrium for $T < 200$ MeV in the order to
secure entropy conservation during hadronization.
This new mechanism of charmonium suppression occurs
due to competition with the yields of $D$ and $D_s$.  $D$ and $D_s$ in effect deplete the pool of available
charmed quark pairs, and fewer hidden charm $c\bar{c}$ mesons are formed. For larger T, right part of
figure~\ref{cc}, the multiplicity of light quarks becomes smaller than equilibrium. This results in an
enhancement of $c\bar{c}$ yield in comparison with chemical equilibrium.

%%%%%%%%%%%%%%%%%%%%%%%%%
%%%%%%%%%%%%%%%%%%%%%%%%%%%%%%

\begin{figure}%[!t]
\centering
\includegraphics[width=9cm,height=9cm]{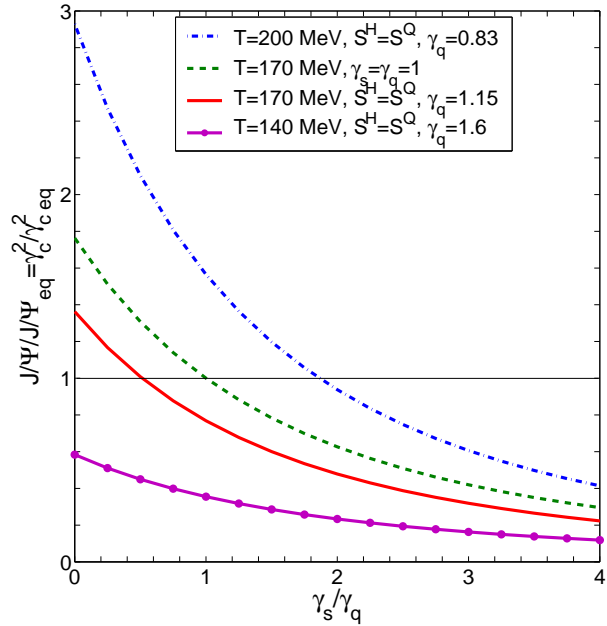}
\caption{\small{(Color on line) Ratio $J\!/\!{\Psi}/J\!/\!\Psi_{eq}=\gamma^2_c/\gamma^2_{c\,eq}$ as a function of $\gamma_s/\gamma_q$ at fixed value of $\gamma_q$ and if required, entropy
conservation.
Shown are: $T=200$ MeV at $\gamma_q=0.83$ (dot-dash line, blue);
$T=170$ MeV at $\gamma_q=1$(dashed line, green) and at  $\gamma_q=1.15$, (solid line, red);  at
$T=140$ MeV, $\gamma_q=1.6$ (solid dotted line, purple)}} \label{jpsrg}
\end{figure}

In  figure~\ref{jpsrg} we quantify a more systematically this
result. We compare the $J\!/\!\Psi$ yield to the chemical
equilibrium yield $(J\!/\Psi)/(J\!/\!\Psi_{eq})$, as a function of
$\gamma_s/\gamma_q$, each line is at a fixed value $\gamma_q$. This
ratio is:
\begin{equation}
\frac{J\!/\!{\Psi}}{J\!/\!{\Psi}_{\rm eq}}=\frac{N_{hid}}{N_{hid\,eq}}=\frac{\gamma_c^2}{\gamma_{c\,{\rm eq}}^2}.
\end{equation}
$(J\!/\!{\Psi})/(J\!/\!{\Psi}_{\rm eq})$  always decreases when $\gamma_s/\gamma_q$ increases, as we expect.
We see that when we have small yields of light $q,s$ quarks, we of course find $J\!/\!\Psi/J\!/\!\Psi_{\rm eq}>1$.
This happens e.g. when hadronization is at $T=200$ MeV. In all other and `reasonable' hadronization
scenarios we find suppression  $(J\!/\!{\Psi})/(J\!/\!{\Psi}_{\rm eq})<1$.

%%%%%%%%%%%%%%%%%%%%%%%%% figure 19
%%%%%%%%%%%%%%%%%%%%%%%%%%%%%%

%%%%%%%%%%%%%%%%%%%%%%%%%%%%%%%%%
\section{Conclusions}\label{concSec}
%%%%%%%%%%%%%%%%%%%%%%%%%%%%%%%%%

We have considered the abundances of heavy flavor hadrons
within the chemical non-equilibrium statistical hadronization model.
We studied how the (relative) yields of strange and non-strange
charmed mesons vary with a QGP produced fixed strangeness content. A
considerable shift of the yield from non-strange $D(B)$ to the
strange $D_s(B_s)$ and similar for strange-charmed baryons is expected for LHC since
$s/S$ may reach value about $0.04$, which upon hadronization means $\gamma_s/\gamma_q \cong 2$.

Another  important consequence of this result is that we find a relative suppression of the
multi-heavy hadrons, including $J/\Psi$, for hadronization temperature lower than $T=200$ MeV,
because strangeness and light quarks are above chemical equilibrium in HG after hadronization. We compared the yields to the expectations based on
chemical equilibrium yields of light and strange quark pairs and
found that the  $J/\Psi$ is suppressed, the mechanism being that
abundantly produced light quark-charmed and strange-charmed hadrons
deplete the pool of available charmed quarks.

\vspace*{.2cm}
\subsubsection*{Acknowledgments}
Work is supported by a grant from: the U.S. Department of Energy  DE-FG02-04ER4131.

%
%%%%%%%%%%%%%%%%%%%%%%%%%%%%%%%%%%%%%%%%%
\vspace*{-0.3cm}
%%%%%%%%%%%%%%%%%%%%%%%%%%%%%%%%%%%%%%%%%
\section{References}
%\begin{references}

\end{document}